\documentclass[reprint,preprintnumbers,amsmath,amssymb,nofootinbib,aps,prx]{revtex4-1}

\usepackage{slashed}
\usepackage{amsmath}
\usepackage{amssymb}
\usepackage{amsthm}
\usepackage{hyperref}
\usepackage{indentfirst}
\usepackage{psfrag}
\usepackage{graphicx}
\usepackage[utf8]{inputenc}

\hypersetup{colorlinks=true, linkcolor=blue, urlcolor=blue, citecolor=blue, linktocpage=true}

\begin{document}
	
	\title{ Unveiling complex vector dark matter by magnetic field }
	
	\author{E. Nugaev$^{1}$}
	\email{emin@ms2.inr.ac.ru}
	\author{A. Shkerin$^{2}$}
	\email{andrey.shkerin@epfl.ch}
	
	\affiliation{$^1$Institute for Nuclear Research of the Russian Academy of Sciences, 60th October Anniversary Prospect, 7a, Moscow, 117312, Russia}
	\affiliation{$^2$Institute of Physics, Ecole Polytechnique F\'ed\'erale de Lausanne,  CH-1015 Lausanne, Switzerland}
	
	\begin{abstract}
		
We discuss the possibility that dark matter is made of a new complex massive vector field with a global $U(1)$-symmetry. The field interacts with the Standard Model via coupling to the gauge field of the hypercharge group. We study classical homogeneous configurations arising in the theory. Some of these configurations include magnetic components of the Standard Model electromagnetic field tensor. This opens a possibility to relate the origin of possible primordial magnetic fields to the cosmological evolution of vector dark matter condensates composed of magnetic bosons.
		
	\end{abstract}
	
	\maketitle

\newcommand{\as}[1]{{\it \color[rgb]{0.000000,0.501961,0.000000}\textbf{as:#1}}}

\section{Introduction}

The nature of dark matter remains elusive. For a long time, the most promising dark matter (DM) candidate was a weakly-interacting massive particle (WIMP) arising naturally in supersymmetric extensions of the Standard Model (SM). Recently, the WIMP-paradigm has come in tension with observations and experiment \cite{Arcadi:2017kky}. On the one hand, refinement in cosmological measurements has challenged the $\Lambda$CDM model of dark matter and dark energy at galactic scales. On the other hand, DM direct detection experiments have failed to find a WIMP particle with standard mass and cross-section. Finally, no signatures of new physics such as supersymmetry has been discovered at the LHC. 

There is a variety of proposals alternative to WIMPs, which explore different mass ranges and/or different forms of coupling between the DM and SM sectors. For example, considering particles with very light masses gives rise to fuzzy dark matter \cite{Hu:2000ke,Hui:2016ltb}. It suggests that DM is in the form of not a dilute gas of particles but rather of a coherent oscillating condensate of them. Both scalar and vector particles were proposed as a constituent of the condensate.\footnote{For recent studies of vector coherent dark matter see, e.g., \cite{AlonsoAlvarez:2019cgw,Nakayama:2019rhg,Nakayama:2020rka}.} From the theory viewpoint, very small masses can be naturally associated with axions and axion-like particles \cite{DiVecchia:2019ejf}. Next, exploring various possible interactions between DM and SM fields leads, e.g., to the Higgs-portal models \cite{Arcadi:2019lka}. Models of millicharged \cite{DeRujula:1989fe,Feng:2009mn} and axion-like particles allow for electromagnetic interaction of DM via the kinetic mixing of dark and visible photons. Another interesting option is to allow the direct coupling of DM particles to the electromagnetic field of the SM. Such a coupling opens new possibilities for direct detection by exploiting electric or magnetic moments of DM particles \cite{Barger:2010gv}.

In this paper, we discuss an extension of the SM by a complex massive vector field $V_\mu$ in which the ideas of coherently oscillating and electromagnetically interacting DM field merge. The synthesis is achieved by introducing a cubic coupling of $V_\mu$ to the gauge field of the hypercharge group of the SM. The new vector is a singlet under the SM symmetry group, and the interaction between the vector and the hypercharge fields is invariant under the gauge $U(1)_Y$-group and also under a global $U(1)$-group associated with $V_\mu$. The theory contains a variety of classical homogeneous solutions of equations of motion -- condensates. We argue that such solutions may represent DM at different stages of evolution of the Universe. The condensed $V$-bosons are never in thermal equilibrium with the cosmic plasma, thus avoiding the constraints for WIMP-like DM candidates.

A complex vector condensate is manifestly not invariant under the global charge conjugation. We will be agnostic about a mechanism generating the asymmetry between $V$-particles and antiparticles. Such a mechanism may operate at energy scales of a UV-completion of the $V$-sector of the model, which is anyway necessary to make the latter renormalizable (see, e.g., \cite{Porrati:2008gv}). Asymmetric conditions for DM can be produced non-perturbatively, e.g., during inflation \cite{Petraki:2013wwa}.

A homogeneous vector condensate breaks spontaneously the rotational symmetry of the FRW metric. Hence, quantum fluctuations of the homogeneous vector field generated during inflation are subject to anisotropy constraints (see, e.g., \cite{Graham:2015rva}).\footnote{We assume that the vector field $V_\mu$ does not dominate the energy budget of the Universe during inflation.} In this paper, we assume that $40-50$ e-foldings before the end of inflation the $V$-sector of the theory was in an isotropic phase so that the condensate has not yet been formed. This assumption is supported by the fact that the energy density of the condensate can increase rapidly as we move backward in time during inflation and quickly pass the cutoff above which the vector field does not longer represent the valid degrees of freedom.

Due to the direct coupling between the vector and the electromagnetic fields at low energies, the model possesses classical solutions involving a homogeneous magnetic field. Such solutions may be relevant for cosmology in view of evidence for large-scale intergalactic magnetic fields, possibly of cosmological origin \cite{Durrer:2013pga}. Primordial magnetic fields may also be necessary to seed the galactic dynamo mechanism \cite{Davis:1999bt}. Existing models of cosmological magnetogenesis mainly operate at inflationary or post-inflationary epochs; see, e.g., \cite{Anber:2006xt,Adshead:2016iae,Cuissa:2018oiw,Kamada:2019uxp} and references therein. An interesting alternative is the late-time magnetogenesis which allows one to bypass constraints from Nucleosynthesis \cite{Choi:2018dqr}. We suggest a new mechanism of the late-time magnetogenesis in which the large-scale homogeneous magnetic field arises as a result of classical instability developed in a vector condensate.\footnote{In our model, the instability arises due to the interaction between the vector and electromagnetic fields and is unrelated to the Jeans instability.} We argue that the instability leads to a rearrangement of the vector field due to which the magnetic field emerges. The time at which the rearrangement occurs depends on the mass of the $V$-field and its coupling constant.\footnote{In this paper, we do not discuss possible cosmological signatures of such a transition between the non-magnetic and magnetic condensates.} As the Universe expands, the interaction energy between the vector and magnetic fields drops quickly, and the two fields become effectively decoupled. 

The paper is organized as follows. In sec. \ref{sec:setup} we introduce the model, write conserved currents and equations of motion for the vector field. In sec. \ref{sec:zoo} we study various classical homogeneous configurations made of vector and magnetic fields. In particular, we discuss their classical stability. Sec. \ref{sec:evolution} is dedicated to the discussion of a possible cosmological scenario involving these configurations. In sec. \ref{sec:concl} we conclude.

\section{Setup}
\label{sec:setup}

\subsection{Lagrangian of the model}

Consider a theory of the complex massive vector field $V_\mu$. The Lagrangian of the theory is invariant under the global $U(1)$-group. This ensures stability of classical solutions against decay into free quanta of the vector field. Denote
\begin{equation}\label{W}
W_{\mu\nu}=V^*_\mu V_\nu-V_\mu V^*_\nu \;.
\end{equation}
The current
\begin{equation}\label{Jt}
J^\mu_T=i\partial_\nu W^{\mu\nu}
\end{equation}
is manifestly conserved due to the antisymmetry of $W_{\mu\nu}$. Thanks to this current, one can construct an invariant interaction term between $V_\mu$ and the SM gauge field, by circumventing the fact that $V_\mu$ itself is a singlet with respect to the SM gauge group. Note that $J^\mu_T$ is not the Noether current corresponding to the global $U(1)$-invariance. The Lagrangian of $V_\mu$ reads as follows:
\begin{equation}\label{L_V}
\mathcal{L}_V=-\frac{1}{2}V^*_{\mu\nu}V^{\mu\nu}+M^2V^*_\mu V^\mu \; ,
\end{equation}
where $V_{\mu\nu}=\partial_\mu V_\nu-\partial_\nu V_\mu$. The contraction of indices is performed with the metric $g_{\mu\nu}$. The interaction between $V_\mu$ and the SM is written as follows:
\begin{equation}\label{L_int}
\mathcal{L}_{\text{int}}=\frac{i\tilde{\gamma}}{2}W_{\mu\nu}B^{\mu\nu} \;,
\end{equation}
where $B^{\mu\nu}$ is the field strength of the hypercharge field. Overall, the theory is determined by the Lagrangian
\begin{equation}\label{L_tot}
\mathcal{L}_{\text{tot}}=\mathcal{L}_{\text{SM}}+\mathcal{L}_V+\mathcal{L}_{\text{int}} \;.
\end{equation}

We will be interested in the low-energy limit of the theory (\ref{L_tot}), characteristic for the Universe after preheating. Hence, we will focus on the electromagnetic part of the interaction term
\begin{equation}\label{L_IntA}
\mathcal{L}_{\rm int} \supset \mathcal{L}_{VA}=\frac{i\gamma}{2}W_{\mu\nu}F^{\mu\nu} \; ,
\end{equation}
where $\gamma=\tilde{\gamma}\cos\theta_W$ and $F^{\mu\nu}$ is the field strength of the electromagnetic field. Due to this term, non-relativistic $V$-particles propagating in a background magnetic field acquire the dipole magnetic moment \cite{Pauli:1941zz,Lee:1962vm}
\begin{equation}\label{mu}
\mu=\dfrac{\gamma}{2M} \; .
\end{equation}
On can, therefore, think of quanta of the $V$-field as magnetic bosons and of a condensate of such quanta as a system of properly aligned magnetic dipoles interacting via a long-range force.\footnote{See \cite{Lahaye_2009} for a review of properties of dipole bosonic quantum gases.}

The energy scale above which the theory must be UV-completed is estimated as \cite{Porrati:2008gv}
\begin{equation}\label{Cutoff}
\Lambda\sim \frac{M}{\sqrt{\gamma}} \;.
\end{equation}
In what follows, we will treat the Lagrangian (\ref{L_tot}) as that of a low-energy effective theory. A particular form of the UV-completion will not be important for our purposes. We will only assume that a successful condensation of $V$-particles into a homogeneous configuration may take place once the energy drops below the cutoff (\ref{Cutoff}). We will see that, depending on phenomenological constraints on $M$ and $\gamma$, the cutoff scale can take values all the way up to the Planck scale.

In the effective field theory approach, one should consider adding to the Lagrangian (\ref{L_tot}) other interaction terms with the dimension not greater than four. Consider first the operator dual to (\ref{L_IntA}):
\begin{equation}\label{L_IntA_dual}
\tilde{\mathcal{L}}_{VA}=\frac{i\gamma}{2} W_{\mu\nu}\tilde{F}^{\mu\nu} \; , ~~~ \tilde{F}^{\mu\nu}=\epsilon^{\mu\nu\rho\sigma}F_{\rho\sigma} \;.
\end{equation}
Adding it would provide $V$-particles with the electric dipole moment, hence the theory would admit homogeneous configurations with a constant electric field. These are not relevant for our purposes, and we discard the term (\ref{L_IntA_dual}) by requiring the theory to be parity-invariant. Other operators are the quartic self-interaction term of the vector field and the coupling term between the vector and the Higgs fields:
\begin{equation}
\propto (V_\mu^*V^\mu)^2 \; , ~~~ \propto H^\dagger H V_\mu^*V^\mu \; .
\end{equation}
These terms are generated by perturbative quantum corrections and are naturally of the order $\tilde{\gamma}^4$ and $(g'\tilde{\gamma})^2$ correspondingly. Hence, as long as $\tilde{\gamma}\ll 1$, we can neglect these operators to the leading order in $\tilde{\gamma}$ in the classical analysis.\footnote{Furthermore, the Higgs and $Z$-bosons provide short-range interactions between $V$-particles, which are inessential in the discussion of classical solutions.}

To summarize, thanks to the complexity of the vector field $V_\mu$, our model features the single vector-photon coupling term (\ref{L_IntA}). It makes the model different from theories of millicharged or axion-like particles.

\subsection{Equations of motion and conserved currents}
\label{ssec:EOMs}

In what follows, we will be interested in energy scales in the Universe which are much below the scale of restoration of the Electroweak symmetry. Hence, we consider the interaction of $V_\mu$ with the electromagnetic field only, eq. (\ref{L_IntA}). Then, the equation of motion for the vector field and the generalization of the Maxwell equations read as follows:
\begin{align}
& \partial_\mu V^{\mu\nu}+M^2 V^\nu+i\gamma F^{\nu\mu}V_\mu=0 \; , \label{EOMs1}\\ 
& \partial_\mu F^{\mu\nu}-i\gamma \partial_\mu W^{\mu\nu}=0 \; . \label{EOMs2}
\end{align}
We note that the Maxwell equations can be resolved with the particular solution
\begin{equation}\label{F(W)}
F_{\mu\nu}=i\gamma W_{\mu\nu}+C\epsilon_{\mu\nu} \; ,
\end{equation}
where $C$ is a constant and $\epsilon_{\mu\nu}$ is the rank-two antisymmetric tensor. $C$ plays the role of an external magnetic field. We consider the corresponding solution in sec. \ref{ssec:magn}, and for other solutions we assume $C=0$.\footnote{A nonzero $C$ would complicate the relation between non-magnetic and magnetic condensates; see sec. \ref{sec:zoo}. Besides, it can play an important role in the analysis of non-homogeneous solutions arising in the theory (\ref{L_tot}). We leave this analysis to future work.} Plugging (\ref{F(W)}) to (\ref{EOMs1}), we obtain the equation containing the vector field only:
\begin{equation}\label{EOM_V}
\partial_\mu V^{\mu\nu}+M^2 V^\nu-\gamma^2W^{\nu\mu}V_\mu+i\gamma C\epsilon^{\nu\mu}V_\mu=0 \; .
\end{equation}
One can think of it as a result of integrating the photon field out. We see that the coupling of $V_\mu$ to $F_{\mu\nu}$ induces the cubic self-interaction of $V_\mu$. 

The (symmetric) energy-momentum tensor corresponding to the free complex vector field is
\begin{equation}\label{EMT_Vfree}
\begin{split}
T_{\mu\nu}^{\rm free}&= V_\mu^{*\alpha}V_{\alpha\nu}+V_\nu^{*\alpha}V_{\alpha\mu}+M^2(V_\mu^*V_\nu+V_\nu^*V_\mu) \\
& -g_{\mu\nu}(-\frac{1}{2}V^*_{\alpha\beta}V^{\alpha\beta}+M^2V^*_\alpha V^\alpha) \; .
\end{split}
\end{equation}
It is conserved on the free equation of motion for $V_\mu$, since $\partial_\mu V^\mu=0$ for the free theory. The interaction term (\ref{L_IntA}) leads to an additional traceless part:
\begin{equation}\label{EMT_int}
T_{\mu\nu}^{\rm int}=i\gamma(F_\mu\;^\rho W_{\nu\rho}+F_\nu\;^\rho W_{\mu\rho})-ig_{\mu\nu}\frac{\gamma}{2}F_{\rho\sigma}W^{\rho\sigma} \; .
\end{equation}
The full energy-momentum tensor of the theory (\ref{L_tot}) associated with the vector field is
\begin{equation}\label{Current_T}
T_{\mu\nu}^{\rm tot}=T_{\mu\nu}^{\rm free}+T_{\mu\nu}^{\rm int} \; .
\end{equation}
It is conserved on the full equations of motion (\ref{EOMs2}), (\ref{EOM_V}). Note that for all solutions studied below, $T_{\mu\nu}^{\rm tot}$ will satisfy the weak energy condition as long as the magnetic field strength does not exceed $\Lambda^2$. This agrees with the fact that above the cutoff the theory (\ref{L_tot}) must be modified.

Finally, the global $U(1)$-symmetry leads to a current
\begin{equation}\label{Current_Q}
J_Q^\mu=i(V^{*\mu\nu}V_\nu-V^{\mu\nu}V^*_\nu) \; ,
\end{equation}
and the corresponding conserved charge $Q$.

\subsection{Symmetric vs asymmetric dark matter}

The theory (\ref{L_tot}) suggests two ways to look for a DM candidate. The first way is to assume that DM is in the form of a dilute gas containing same portions of non-relativistic particles and antiparticles. From eq. (\ref{EMT_Vfree}) one then concludes that such a gas behaves like pressureless dust. In the scenario of symmetric DM, the constraint on the coupling constant $\gamma$ comes from the annihilation process of $V$ and anti-$V$ to two photons. On dimensional grounds, the cross-section of the process is estimated as $\sigma\sim\gamma^4 M^{-2}$. Measurements of the diffuse gamma-ray background then give (see, e.g., \cite{Gaskins:2016cha})\footnote{We took the conservative estimate $\langle\sigma v\rangle< 10^{-25}$cm$^3$s$^{-1}$ with $v\sim 10^{-3}$. }
\begin{equation}\label{BoundSym}
\frac{1}{\gamma^2}\frac{M}{\rm GeV}>10^{-6} \;.
\end{equation}
Note that this constraint is derived under the assumption of a comparable number of particles and antiparticles. If one violates the assumption by introducing a strong chemical potential, then, in general, the bound on the coupling constant is lost. However, the theory (\ref{L_tot}) allows one to derive a different kind of bound in the asymmetric case, by assuming that a significant portion of DM is not in the form of the dilute gas, but rather in the form of a coherent condensate of $V$-particles. The condensate is manifestly non-invariant under the charge conjugation, hence the condition (\ref{BoundSym}) is not applicable. The vector condensate is the second DM candidate in the theory (\ref{L_tot}), and in the rest of the paper we focus on exploring its properties.

\section{Zoo of homogeneous condensates}
\label{sec:zoo}

\subsection{Non-magnetic condensate}
\label{ssec:vec}

We begin to study classical homogeneous configurations arising in the theory (\ref{L_tot}), which involve vector and/or magnetic fields. In this section we study solutions in flat spacetime, and in the next section we discuss their generalization to the FRW metric.

It is well-known that due to the long-range Coulomb force gauge vector fields do not permit homogeneous solutions. This is not the case for the global massive vector field. Choose the following stationary Ansatz for $V_\mu$:
\begin{equation}\label{Sol_V_free}
V^0=V^3=0 \; , ~~~ V^1=V^2=ve^{-iMt}
\end{equation}
with $v$ a constant. It satisfies the condition $\partial_\mu V^\mu=0$. The phases of the components of $V_\mu$ are synchronized and, as a result, $W_{\mu\nu}=0$. Hence, the Ansatz (\ref{Sol_V_free}) represents the non-magnetic condensate. It satisfies the equation of motion (\ref{EOM_V}) at any $v$. The energy density and the global charge density of the condensate are given by
\begin{equation}\label{VC_EQ}
\rho_E=T_{00}^{\rm free}=2v^2M^2 \; , ~~~ \rho_Q=J^0_Q=2v^2M \; .
\end{equation}
Note that
\begin{equation}\label{EMQ_v}
\rho_E=M\rho_Q \; ,
\end{equation}
hence the condensate is a coherent collection of free $V$-particles. This is expected, since in the absence of the classical electromagnetic field, $V_\mu$ is free. Below we restrict our consideration to the region $\rho_E\ll\Lambda^4$.

Let us study linear classical stability of the solutions (\ref{Sol_V_free}). Since the background is homogeneous, its perturbations can be decomposed into plane waves. The condensate is stable under perturbations of $V^0$ and $V^3$. For the other two components we adopt the following perturbation Ansatz:
\begin{align}
& \delta V^1=e^{-iMt}e^{\alpha t}(v_1 e^{i\textbf{k}\textbf{x}}+u_1 e^{-i\textbf{k}\textbf{x}})  \; , \label{Pert_V_free1} \\
& \delta V^2=e^{-iMt}e^{\alpha t}(v_2 e^{i\textbf{k}\textbf{x}}+u_2 e^{-i\textbf{k}\textbf{x}}) \; ,	\label{Pert_V_free2}
\end{align}
where $v_1$, $u_1$, $v_2$, $u_2$ are complex numbers and $\alpha$ is real. Note that $\delta W^{12}\neq 0$, hence the perturbations involve the electromagnetic field. The linearized equations of motion are reduced to
\begin{align}
& ((\alpha-iM)^2+\textbf{k}^2+M^2)v_1+2\gamma^2v^2(u^*_1-v_1)=0 \; ,\\
& ((\alpha-iM)^2+\textbf{k}^2+M^2)u_1+2\gamma^2v^2(v^*_1-u_1)=0 \; ,
\end{align}
and $v_2=-v_1$, $u_2=-u_1$. This system of homogeneous equations has a solution whenever
\begin{equation}\label{VC_alpha}
\alpha^2+\textbf{k}^2=2v^2\gamma^2 ~~~ \text{or} ~~~ \alpha^2+\textbf{k}^2=4v^2\gamma^2 \; .
\end{equation}
We see that the maximum value of the decay constant $\alpha$ is $2v\gamma$, and the corresponding decay mode is homogeneous. The modes with the non-zero momentum $\textbf{k}$ up to $|\textbf{k}|=2v\gamma$ are also present in the spectrum, see the left panel of fig. \ref{fig:Alpha}. We will make use of the fact that the non-magnetic condensate prefers to decay via long-wavelength modes. The lifetime of the condensate is estimated as
\begin{equation}\label{LifeTimeFlat_VC}
\tau_{VC}\sim \dfrac{1}{v\gamma} \;.
\end{equation}

\subsection{Magnetic field}
\label{ssec:magn}

For completeness, let us consider the case when a homogeneous magnetic field is not sourced by the vector field $V_\mu$. This corresponds to choosing a non-zero constant $C\equiv -B$ in eq. (\ref{F(W)}), with $B$ the magnetic field strength. The non-zero components of the electromagnetic field tensor are
\begin{equation}\label{MF}
F_{ij}=-B\epsilon_{ij} \; , ~~~ (i,j)=(1,2),(2,3),(1,3) \; .
\end{equation}
One can study classical stability of the solution under perturbations of the vector field. For example, if we take $(i,j)=(1,2)$, then the relevant perturbation Ansatz is
\begin{equation}
\delta V^1=e^{\alpha t}e^{i\textbf{k}\textbf{x}}v_1 \; , ~~~ \delta V^2=e^{\alpha t}e^{i\textbf{k}\textbf{x}}v_2 \; ,
\end{equation}
and $\delta V^0=\delta V^3=0$. The linearized equations of motion can be solved for $v_1$, $v_2$ and a real $\alpha$ provided that
\begin{equation}
B^2\gamma^2>M^2(M^2+\textbf{k}^2) \; ,
\end{equation}
i.e. if
\begin{equation}
|B|>\Lambda^2 \; .
\end{equation}
Thus, magnetic fields with the strength above the cutoff are unstable, which agrees with the discussion of the energy-momentum tensor in sec. \ref{ssec:EOMs}. Below we require the cutoff to exceed the Schwinger value $\sim 1\:$MeV. Note finally that the theory admits linear superpositions of the orthogonal vector and magnetic fields. For example, one can take $(i,j)=(1,2)$ in eq. (\ref{MF}) and supplement it with the 3rd component of $V_\mu$, $V^3=Me^{-iMt}$.

\subsection{Magnetic condensate}

\begin{figure*}[t]
	\begin{center}
		\begin{minipage}[h]{0.45\linewidth}
			\center{\includegraphics[scale=0.7]{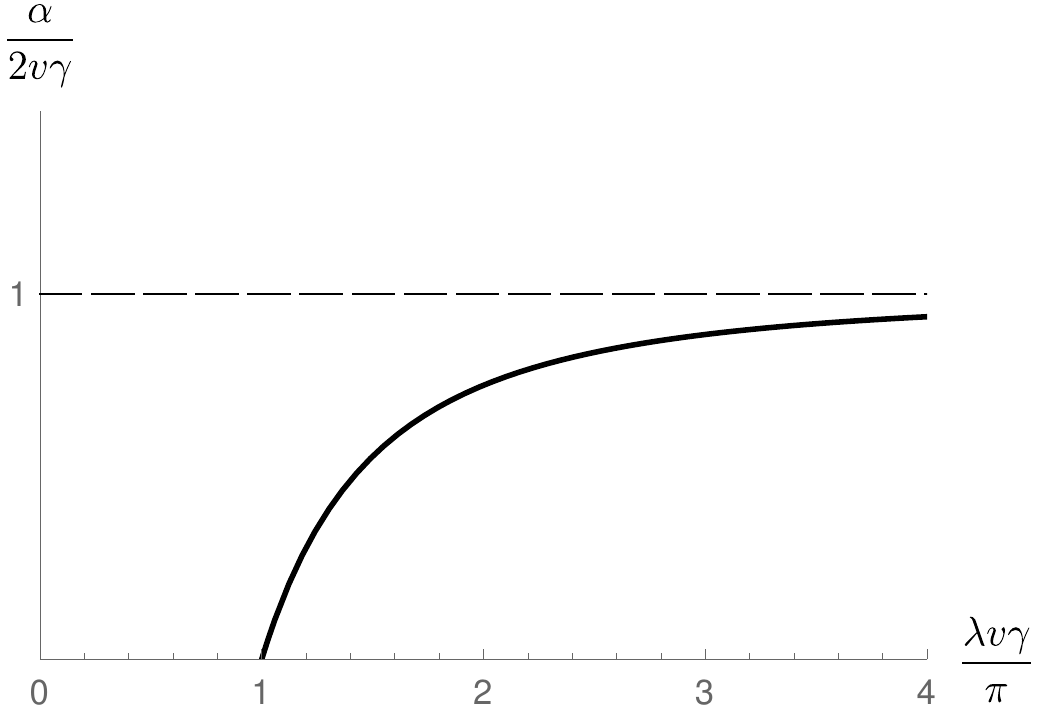} \\ (a)}
		\end{minipage}
		\begin{minipage}[h]{0.45\linewidth}
			\center{\includegraphics[scale=0.7]{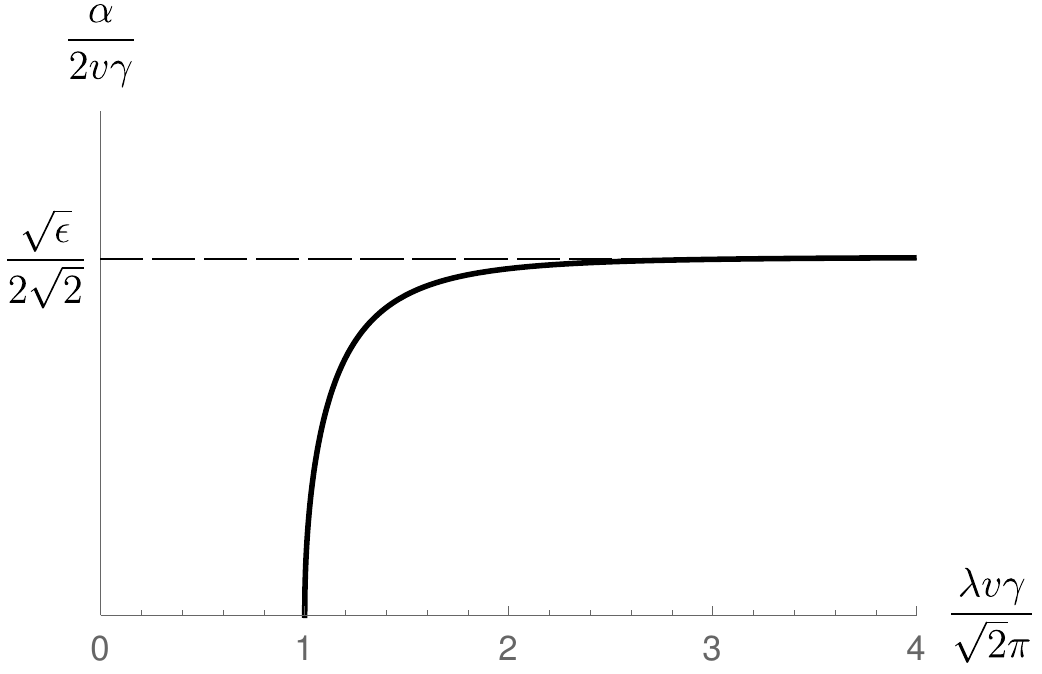} \\ (b)}
		\end{minipage}
		\caption{Decay constant $\alpha$ as a function of the wavelength $\lambda$ of the decay mode, for the non-magnetic (a) and magnetic (b) condensates. In the non-magnetic case, the dominant decay mode is presented.}
		\label{fig:Alpha}
	\end{center}
\end{figure*}

According to eq. (\ref{F(W)}), the magnetic field is induced if $W_{\mu\nu}\neq 0$. To achieve this, we introduce a relative phase between the components of the vector field. Take the following Ansatz for $V_\mu$:
\begin{equation}\label{MC_ansatz}
V^0=V^3=0 \; , ~~~ V^1=ve^{-i\omega t} \; , ~~~ V^2=V^1 e^{i\varphi}
\end{equation}
with $v,\omega,\varphi$ constants and $0\leqslant\omega<M$. Plugging this to the equation of motion (\ref{EOM_V}) and taking $C=0$ gives
\begin{equation}\label{EOM_mc}
(M^2-\omega^2)v+2ie^{i\varphi}\sin\varphi\gamma^2v^3 =0 \; .
\end{equation}
Without loss of generality, we require the magnitude $v$ of the condensate to be real (it can always be made real by applying a global $U(1)$-transformation). This fixes the phase $\varphi=\pm\frac{\pi}{2}$. Next, introduce the parameter
\begin{equation}\label{Epsilon}
\epsilon=\dfrac{M^2-\omega^2}{M^2} \; .
\end{equation}
We will see shortly that $\epsilon$ measures the difference between non-magnetic and magnetic condensates of the same charge density. From eqs. (\ref{F(W)}) and (\ref{EOM_V}) we have
\begin{equation}\label{MC}
v^2=\dfrac{\epsilon M^2}{2\gamma^2} \; , ~~~ B_1=B_2=0 \; , ~~ B_3\equiv B=\pm\dfrac{\epsilon M^2}{\gamma} \; .
\end{equation}
The sign of $B$ is determined by the sign of $\varphi$. Take, for example, $\varphi=\frac{\pi}{2}$, then $B>0$. 

Eqs. (\ref{MC_ansatz}), (\ref{MC}) represent the magnetic condensate we look for. Note again that the magnetic field arises due to the relative phase between the rotating components of the vector field or, in other words, due to a certain arrangement of magnetic bosons constituting the condensate. The arrangement is optimal if the energy density of the configuration is minimal, and this is achieved when the phase shift is $\pm\frac{\pi}{2}$.

Let us compute the charge and energy densities of the magnetic condensate. The former is given by
\begin{equation}\label{Q_mc}
\rho_Q=\dfrac{2\omega\epsilon M^2}{\gamma^2} \; ,
\end{equation}
and the latter is
\begin{equation}\label{E_mc}
\rho_E =M\rho_Q+\rho_{\text{EM}}+\rho_{\text{int}} = \frac{\epsilon(4-3\epsilon)M^4}{2\gamma^2} \;.
\end{equation}
Here
\begin{equation}
\rho_{\text{EM}}=\frac{B^2}{2} \;, ~~~ \rho_{\text{int}}=-\frac{M^4\epsilon^2}{\gamma^2}
\end{equation}
are the energy density of the magnetic field and of the interaction between the vector and magnetic fields, correspondingly. Taking derivatives with respect to $\omega$, we observe that
\begin{equation}
\dfrac{d\rho_E}{d\omega}=\omega\dfrac{d\rho_Q}{d\omega} \; .
\end{equation}
This relation is known to hold for many types of non-topological soliton solutions; see, e.g., \cite{Nugaev:2019vru}.

In what follows, we will work in the regime of small $\epsilon$. It corresponds to the non-relativistic limit $\omega\approx M$. From eqs. (\ref{Cutoff}), (\ref{E_mc}) we see that in this regime $\rho_E\ll\Lambda^4$. Furthermore,
\begin{equation}\label{EQ_mc}
\rho_E\approx M\rho_Q-\frac{\epsilon^2M^4}{2\gamma^2}\;, ~~~ \epsilon\ll 1\;,
\end{equation}
from which we conclude that the particles forming the magnetic condensate are in a bound state, and the free particle limit is reproduced at $\epsilon\rightarrow 0$. Thus, although the magnetic field $\textbf{B}$ gives the positive contribution to $\rho_E$, the interaction energy between the fields $V_\mu$ and $\textbf{B}$ is negative and such that $\rho_E<M\rho_Q$. In other words, the magnetic condensate is more energetically favorable that the non-magnetic one (cf. eq. (\ref{EMQ_v})).\footnote{This conclusion is valid only at $\epsilon\ll 1$.} 

Let us study linear classical stability of the magnetic condensate with respect to perturbations of the vector field. Perturbations of $V^0$ and $V^3$ do not affect the solution. For the other two components we adopt the following perturbation Ansatz (cf. eqs. (\ref{Pert_V_free1}), (\ref{Pert_V_free2})):
\begin{align}
& \delta V^1=e^{-i\omega t}e^{\alpha t}(v_1 e^{i\textbf{k}\textbf{x}}+u_1 e^{-i\textbf{k}\textbf{x}}) \; , \label{Pert_mc1} \\
& \delta V^2=e^{-i\omega t}e^{\alpha t}(v_2 e^{i\textbf{k}\textbf{x}}+u_2 e^{-i\textbf{k}\textbf{x}}) \; , \label{Pert_mc2}
\end{align}
where $v_1$, $v_2$, $u_1$, $u_2$ are complex numbers and $\alpha$ is real. The linearized equations of motion become
\begin{align}
& ((\alpha-i\omega)^2+\textbf{k}^2+M^2)v_1-2\gamma^2v^2(v_1+u^*_1)=0 \; , \\
& ((\alpha-i\omega)^2+\textbf{k}^2+M^2)u_1-2\gamma^2v^2(u_1+v^*_1)=0 \; ,
\end{align}
and $v_2=iv_1$, $u_2=iu_1$. This system of homogeneous equations has a solution if
\begin{equation}
\alpha^4+2\alpha^2(\textbf{k}^2+2(1-\epsilon)M^2)+\textbf{k}^4-\epsilon^2M^4=0 \; ,
\end{equation}
where we used eq. (\ref{MC}). When $\epsilon\ll 1$, the maximum value of the decay constant $\alpha$ is $\alpha\approx\frac{M\epsilon}{2}$, and the corresponding decay mode is homogeneous. There are also decay modes with the non-zero momentum $\textbf{k}$, up to $|\textbf{k}|=M\sqrt{\epsilon}$, see the right panel of fig. \ref{fig:Alpha}. The lifetime of the magnetic condensate is estimated as
\begin{equation}\label{LifeTimeMCFlat}
\tau_{MC}\sim \dfrac{1}{M\epsilon} \; .
\end{equation}

\subsection{Transition between condensates}

As was mentioned above, at a given charge density, the energy density of the magnetic condensate at $\epsilon\ll 1$ is lower than that of the non-magnetic condensate. Given this and the fact that the non-magnetic condensate prefers to decay via long-wavelength modes, it is reasonable to suggest that the configuration (\ref{MC}) results from the classical evolution of the vector configuration (\ref{Sol_V_free}), which essentially amounts to developing the phase shift between $V^1$ and $V^2$. Equating $\rho_Q$ in eqs. (\ref{VC_EQ}) and (\ref{Q_mc}),\footnote{The magnetic condensate retains the same charge density $\rho_Q$, if the excess of energy is carried away by particle-antiparticle pairs. Otherwise, $\rho_Q$ changes by the amount $\mathcal{O}(\epsilon^2)$, and in the non-relativistic limit we can neglect the difference. For the same reason, we can neglect the number density of the allegedly produced $V$ and anti-$V$ particles.} we express the magnitude $v$ of the condensate through $\epsilon$. Then eq. (\ref{LifeTimeFlat_VC}) gives
\begin{equation}\label{LifeTimeVC2}
\tau_{VC}\sim\frac{1}{M\sqrt{\epsilon}} \;.
\end{equation}
Comparing with eq. (\ref{LifeTimeMCFlat}), we see that the magnetic condensate lives parametrically longer than the non-magnetic one. Hence, the former can indeed be an intermediate step of the evolution of the latter. In the next section, we will work under this conjecture and assign to the magnetic field emerging this way the role of the primordial magnetic field which may survive until present time in the intergalactic medium.

The form (\ref{Pert_V_free1}), (\ref{Pert_V_free2}) of the perturbation suggests that the 0th and 3rd components of $V_\mu$ remain undisturbed during the transition. The resulting magnetic field can have two different orientations depending on the sign of the phase shift, see eqs. (\ref{MC}). It makes sense, therefore, to speak about a correlation length $\ell_{\rm corr}$ of the magnetic field as an average distance at which the latter points at the same direction. The correlation length is estimated as a characteristic wavelength of the decay mode, hence
\begin{equation}\label{Lcorr}
\ell_{\rm corr}\sim\tau_{VC} \;.
\end{equation}

\section{Cosmological evolution of condensates}
\label{sec:evolution}

\subsection{Condensates in the expanding Universe}
\label{ssec:evcond}

In this section we study how the homogeneous solutions considered above are embedded into an external FRW metric
\begin{equation}
ds^2=dt^2-a(t)^2dl^2 \;.
\end{equation}
We will focus on the late-time dynamics of the $V$-field, by assuming that the Hubble rate $H$ is much below the mass of the boson $M$. We will not consider a possible evolution of the non-magnetic condensate preceding the time when $H\ll M$. As discussed in sec. \ref{sec:setup}, this evolution may involve physics above the cutoff scale $\Lambda$, of which we are not aware. Under the assumption that the non-magnetic condensate has already been formed at a time $\tilde{t}$, its magnitude at subsequent times scales as (see appendix \ref{app:CondInFRW} for details)
\begin{equation}\label{VinFRW1}
V_{1,2}(t)=\tilde{V}_{1,2}\sqrt{\dfrac{\tilde{a}}{a(t)}}e^{-iMt} \; ,
\end{equation}
where $\tilde{a}=a(\tilde{t})$. We see that $V_\mu\sim a^{-1/2}$, hence $V^\mu\sim a^{-5/2}$ and $V^*_\mu V^\mu\sim a^{-3}$. Thus, the non-magnetic condensate behaves like pressureless dust. Straightforward calculation confirms that
\begin{equation}
\rho_E(t)=\tilde{\rho}_{E}\left(\dfrac{\tilde{a}}{a(t)}\right)^3 \; .
\end{equation}

As discussed in sec. \ref{ssec:vec}, the non-magnetic condensate is classically unstable due to the self-interaction of $V_\mu$ induced by the photon field. We argued that a possible outcome of the decay is a magnetic condensate of a (slightly) lower energy density. Let $\bar{t}$ be the moment of formation of the magnetic condensate. It is convenient to parametrize its magnitude by the parameter $\epsilon$. The latter can be related to the magnitude $\bar{v}$ of the non-magnetic condensate at $t=\bar{t}$ by using the conservation of charge. Namely, from eqs. (\ref{VC_EQ}), (\ref{Q_mc}) we have
\begin{equation}\label{epsV}
\epsilon=\frac{\gamma^2\bar{v}^2}{M^2} \;.
\end{equation}
At $t>\bar{t}$, the vector components of the magnetic condensate decrease as (see appendix \ref{app:CondInFRW} for details)
\begin{equation}\label{VC_V_FRW}
V_{1,2}(t)=\bar{V}_{1,2}\sqrt{\frac{\bar{a}}{a(t)}}e^{-i\omega t} \;,
\end{equation}
where
\begin{equation}\label{VC_V_FRW2}
\bar{V}_1=\frac{\sqrt{\epsilon}M}{\sqrt{2}\gamma} \; , ~~~ \bar{V}_2=\pm i\bar{V}_1 \;.
\end{equation}
Again, we see that $V_\mu^* V^\mu\sim a^{-3}$. Since the dominant contribution to the energy density is provided by the vector field, this shows that the mangetic condensate also behaves like a dust, so that
\begin{equation}\label{RhoEFRW}
\rho_{E,Q}(t)=\bar{\rho}_{E,Q} \left(\dfrac{\bar{a}}{a(t)}\right)^3 \; .
\end{equation}
On the other hand, the magnetic field strength at $t>\bar{t}$ is given by 
\begin{equation}\label{BFRW}
B_3(t)=\bar{B}_3\frac{\bar{a}}{a(t)} \; , ~~~ \bar{B}_3=\pm\frac{\epsilon M^2}{\gamma} \;.
\end{equation}
Hence, $B_3\sim a^{-1}$, as expected for the covariant component of magnetic field  (see, e.g., \cite{Subramanian:2015lua}). Consequently, $B^3\sim a^{-3}$ and $B_i B^i\sim a^{-4}$, as it should. From eqs. (\ref{LifeTimeVC2}) and (\ref{Lcorr}) we see that the magnetic field is correlated at distances
\begin{equation}\label{lCorrFRW}
\ell_{\rm corr}(t)\sim\frac{1}{M\sqrt{\epsilon}}\frac{a(t)}{\bar{a}} \;.
\end{equation}
Finally, from eqs. (\ref{EMT_int}), (\ref{VC_V_FRW}) we obtain that the energy density of interaction between $V_\mu$ and $\textbf{B}$ scales as
\begin{equation}
\rho_{E,\rm{int}}(t)=\bar{\rho}_{E,\rm{int}}\left(\frac{\bar{a}}{a(t)}\right)^6 \;.
\end{equation}
It means that soon after formation, $\textbf{B}$ becomes effectively decoupled from $V_\mu$, and the two fields can evolve independently.

The vector field constituting the magnetic condensate is subject to the Jeans instability. Self-interaction does not affect the growth of gravitational perturbations as long as the gravitational energy stored in these perturbations exceeds the energy due to the self-interaction:
\begin{equation}
\begin{split}
E_{\text{grav}} & \sim\dfrac{\delta\rho_E}{\rho_E}\rho_E\\
& \sim\phi\dfrac{\epsilon M^4}{\gamma^2}\left(\dfrac{\bar{a}}{a}\right)^3>\dfrac{M^4\epsilon^2}{\gamma^2}\left(\dfrac{\bar{a}}{a}\right)^6 \;,
\end{split}
\end{equation}
where $\phi$ is the gravitational potential, or
\begin{equation}
\phi>\epsilon\left(\dfrac{\bar{a}}{a}\right)^3 \;.
\end{equation}
As we will see shortly, phenomenology requires $\epsilon$ to be very small, and the last condition can be fulfilled easily.

\subsection{Dark matter abundance and magnetogenesis}
\label{ssec:DMMG}

Our goal is to demonstrate how the vector part of the magnetic condensate can be responsible for the dark matter abundance at the same time as its magnetic part is responsible for the primordial magnetic field. Here we only intent to provide an example of the calculation, leaving the detailed analysis of the reliability of the suggested mechanism to future work. 

To make our discussion quantitative, we will employ the following cosmological bounds on the magnetic field strength. First, to explain observed galactic fields, the lower bound on the primordial magnetic field \cite{Davis:1999bt} was suggested at the time $z_*\sim 10$ of formation of first galaxies \cite{Barkana:2000fd}:
\begin{equation}\label{B01}
B_*\gtrsim 10^{-30} \:\text{G} \;.
\end{equation}
On the other hand, the growing evidence for large-scale intergalactic magnetic fields inferring from gamma-ray observations suggests that \cite{Durrer:2013pga}
\begin{equation}\label{B02}
B_0\gtrsim 10^{-19}\:\rm{G} \;.
\end{equation}
Finally, as an upper bound, we will use the constraint derived from CMB anisotropy measurements \cite{Pshirkov:2015tua}:\footnote{Big Bang Nucleosynthesis also provides a bound on the magnetic field present at $z_{\rm BBN}\sim 10^9$; for today's field strength it gives $B_0\lesssim 1.5\:\mu$G \cite{Kawasaki_2012}, which is weaker than the CMB bound (\ref{B03}). }
\begin{equation}\label{B03}
B_0\lesssim 10^{-9}\:\rm{G} \;.
\end{equation}
In applying the conditions (\ref{B02}), (\ref{B03}), we assume that the primordial magnetic field survived until present time in the intergalactic medium where its strength were simply rescaling from the moment of formation. 

For simplicity, assume that the $V$-field of the condensate represents a significant portion of dark matter, 
\begin{equation}\label{DM1}
\Omega_V\sim\Omega_{\rm DM} \;. 
\end{equation}
Next, we estimate the time of formation of the magnetic condensate as $\bar{t}\sim\tau_{VC}$ and focus on the regime in which the magnetic field emerges in the radiation-domination epoch:
\begin{equation}\label{t1}
\tau_{VC}<t_{\rm eq} \; ,
\end{equation}
where $t_{\rm eq}$ refers to the moment of matter-radiation equality. We will see that the latter assumption is consistent with the conditions (\ref{B01})--(\ref{DM1}). By using eq. (\ref{LifeTimeVC2}), the bound (\ref{t1}) is rewritten as
\begin{equation}\label{Cond1}
\sqrt{\epsilon}\frac{M}{\rm GeV}>10^{-36} \;.
\end{equation}

Let us start by evaluating the vector field abundance. From eqs. (\ref{E_mc}), (\ref{RhoEFRW}) we have
\begin{equation}
\rho_E(t_{\rm eq})\sim\bar{\rho}_E\left(\frac{\tau_{VC}}{t_{\rm eq}}\right)^{3/2}\sim\frac{\epsilon^{1/4}M^{5/2}}{\gamma^2}\frac{T_{\rm eq}^3}{M_P^{*3/2}} \;.
\end{equation}
Requiring $\Omega_V\sim \rho_E(t_{\rm eq})/\rho_\gamma(t_{\rm eq})\sim 1$, where $\rho_\gamma$ is the radiation energy density, and plugging in numbers, we obtain
\begin{equation}\label{Cond2}
\frac{\sqrt{\epsilon}}{\gamma^4}\left(\frac{M}{\rm GeV}\right)^5\sim 10^{36} \;.
\end{equation}
Next, let us evaluate the magnetic field strength at the moment $t_*$:
\begin{equation}
B_*\sim\bar{B}\frac{\tau_{VC}z_*^2}{t_{\rm eq}z_{\rm eq}^2}\sim\frac{\sqrt{\epsilon}M}{\gamma}\frac{T_{\rm eq}^2z_*^2}{M_P^*z_{\rm eq}^2} \;,
\end{equation}
where we used eq. (\ref{BFRW}). Plugging in numbers and implementing the conditions (\ref{B01}) and (\ref{B03}), we arrive at
\begin{equation}\label{Cond3}
\frac{\sqrt{\epsilon}}{\gamma}\frac{M}{\rm GeV}=10^{-7}b \; ,
\end{equation}
where the parameter $b$ can vary from $\mathcal{O}(1)$ to $\mathcal{O}(10^{23})$. The conditions (\ref{Cond1}), (\ref{Cond2}) and (\ref{Cond3}) are consistent with each other provided that $\gamma>10^{-29}b^{-1}$. 

The requirements (\ref{Cond2}), (\ref{Cond3}) fix two out of the three parameters of the solution. It is convenient to rewrite them as one condition on the mass and the coupling constant:
\begin{equation}\label{Cond4}
\frac{1}{\gamma^3}\left(\frac{M}{\rm GeV}\right)^4=10^{43}b^{-1} \;.
\end{equation}
For example, let us take the lower bound (\ref{B01}) on the magnetic field strength, $b=1$, and $M=1\:$KeV. Then we have $\gamma\sim 10^{-23}$. Increasing the magnetic fields at a given $M$ amounts to decreasing $\gamma$. We conclude that our mechanism allows one to probe very small values of the coupling constant. This follows from the non-perturbative nature of the condensates which results in the inverse dependence of their magnitudes on $\gamma$. Note that the occupation number of $V$-particles in the condensate can be very small at the values of $M$ above the ones characteristic for fuzzy dark matter. This does not prevent the configuration from being coherent and described by the solution of classical equations of motion, as soon as $\gamma$ is small enough (see, e.g., \cite{Dalfovo:1999zz}).\footnote{The occupation number is large at the time $t_{\rm init}$ of formation of the non-magnetic condensate provided that $\rho_{E,\rm{init}}\sim\Lambda^4$.} At the given values of $b$ and $M$, one also obtains $\epsilon\sim 10^{-47}$ and $\Lambda\sim 100\:$TeV. The magnetic condensate emerges when the Universe is $\tau_{VC}\approx 1.5$ days old. Finally, from eq. (\ref{lCorrFRW}) we deduce the correlation length of the magnetic field $\ell_{\rm corr}(t_*)\approx 3\:$kpc.

\begin{figure}[t]
	\begin{center}
			\center{\includegraphics[scale=0.65]{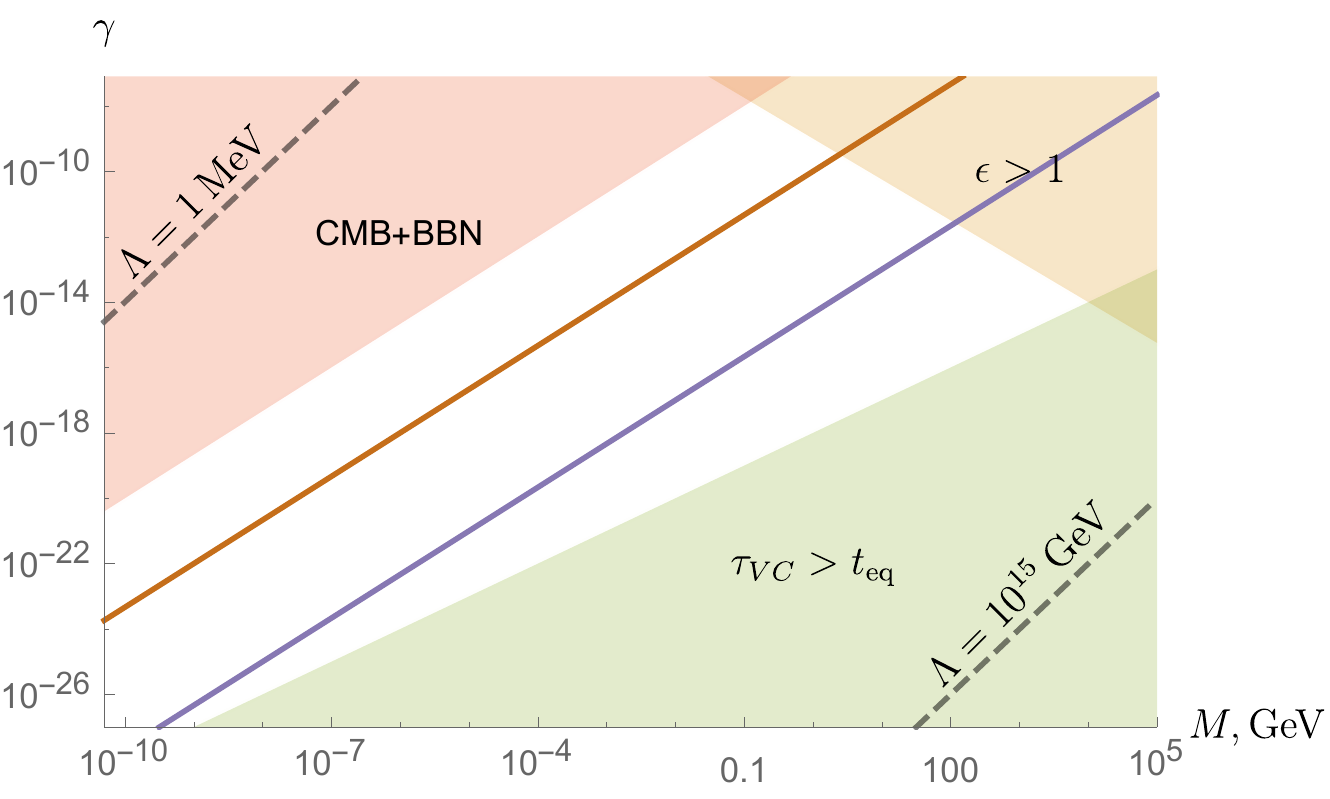}}
		\caption{The range of admissible values of $M$ and $\gamma$, constrained by the CMB bound (\ref{B03}) from the top (see also footnote 10), by the requirement of the condensate to be non-relativistic from the right, and by the assumption that the magnetic condensate forms in the radiation-domination epoch from the bottom. The dashed lines enclose the region of cutoff values from the scale $\mathcal{O}(1)\:$MeV to the typical inflationary scale $\mathcal{O}(10^{15})\:$GeV. The lower solid line corresponds to the magnetic field strength $B_*\sim 10^{-30}$ G at $z_*\sim 10$, the upper solid line corresponds to today's strength $B_0\sim 10^{-19}$ G. We work under the condition (\ref{DM1}).}
		\label{fig:MainPlot}
	\end{center}
\end{figure}

Let us comment on the assumption (\ref{t1}) we made in deriving eq. (\ref{Cond4}). The late-time formation of the magnetic condensate is also possible, but we find it less reliable, since, in view of eq. (\ref{LifeTimeVC2}), it would lead to much smaller values of $\gamma$ than in the example above. As a result, unless $M$ is carefully tuned, this would make the cutoff $\Lambda$ exceed the Planck scale.

In fig. \ref{fig:MainPlot} we summarize the above analysis and plot the window of admissible values of $M$ and $\gamma$. The two benchmark lines correspond to the magnetic field strength saturating the bounds (\ref{B01}) and (\ref{B02}). Note again that we work under the condition that most of the dark matter is condensed in the coherent homogeneous configuration. Relaxing this condition or allowing for other sources of cosmological magnetic fields enhances the allowable region of parameters. It is also interesting to note that the bound (\ref{BoundSym}), which is applicable for the symmetric dark matter, probes an entirely different parameter range. Same is true for the constraints from dispersion measure observations which are typical for millicharged particles. Indeed, adopting the results of Ref. \cite{Caputo:2019tms}, we obtain 
\begin{equation}
\dfrac{1}{\gamma}\dfrac{M}{\text{GeV}}\gtrsim 0.1\sqrt{\dfrac{\rho_E}{0.3\:\text{GeV}/\text{cm}^3}} \;.
\end{equation}
Clearly, this does not interfere with eq. (\ref{Cond1}) or (\ref{Cond4}).

\section{Conclusion}
\label{sec:concl}

In this paper we introduced an extension of the Standard Model by a complex massive vector field $V_\mu$ with a global $U(1)$-invariance. The model features the direct coupling (\ref{L_IntA}) between the vector and photon fields, which allows us to treat $V$-particles as feebly interacting magnetic dipoles. We studied classical homogeneous solutions arising in the model and discussed their cosmological implications. The solutions represent condensates of $V$-particles; some of them support internal magnetic field. We presented a proof of concept that such a condensate can simultaneously account for dark matter and a possible cosmological magnetic field. The vector-photon interaction opens new possibilities for observational tests involving large magnetic fields, e.g., in magnetars.

The authors thank D. Gorbunov, M. Shaposhnikov and S. Troitsky for useful discussions. The work of A.S. was supported by the Swiss National Science Foundation Excellence grant 200020B\underline{ }182864.

\appendix

\section{Evolution of condensates in the FRW metric}
\label{app:CondInFRW}

It is convenient to discuss homogeneous solutions arising in the theory (\ref{L_tot}) in the conformal time:
\begin{equation}\label{MetricAnsatz}
ds^2=a(\eta)^2(d\eta^2-dl^2) \; .
\end{equation}
The equation of motion for the vector field with the induced self-interaction becomes (cf. eq. (\ref{EOM_V}))
\begin{equation}\label{EoM11}
\eta^{\mu\rho}\partial_\mu V_{\rho\nu}+M^2a(\eta)^2V_\nu-\gamma^2\eta^{\rho\sigma}W_{\nu\rho}V_\sigma=0 \; .
\end{equation}
Consider first the non-magnetic condensate. We take the following Ansatz:
\begin{equation}\label{vEta}
V_1=V_2=v(\eta)e^{-i\Omega(\eta)}
\end{equation}
and $V_0=V_3=0$. Here $v$ and $\Omega$ are real functions of the conformal time. Separating real and imaginary components gives
\begin{equation}\label{TwoEqs}
\ddot{v}-v\dot{\Omega}^2+vM^2a^2=0 \; , ~~~ \ddot{\Omega}v+2\dot{\Omega}\dot{v}=0 \; ,
\end{equation}
where dot means derivative with respect to $\eta$. The second equation can be integrated to yield
\begin{equation}\label{v(omega)}
v=v_*(\dot{\Omega})^{-1/2} \; .
\end{equation}
Plugging this to the first equation gives
\begin{equation}\label{EqForOmega}
-\dfrac{1}{2}\dddot{\Omega}\dot{\Omega}+\dfrac{3}{4}\ddot{\Omega}^2-\dot{\Omega}^4+M^2a^2\dot{\Omega}^2=0 \; .
\end{equation}
In the regime $M\gg H$, let us neglect the first two terms of eq. (\ref{EqForOmega}). Then, the approximate solution is
\begin{equation}\label{solOmega}
\dot{\Omega}=Ma \; ,
\end{equation}
and from eqs. (\ref{vEta}), (\ref{v(omega)}) and (\ref{solOmega}) we have
\begin{equation}\label{VinFRW}
V_{1,2}(t)=v_*\sqrt{\dfrac{a_*}{a(t)}}e^{-iMt} \; ,
\end{equation}
which reproduces eq. (\ref{VinFRW1}).

Let us check the approximation we made in neglecting the first two terms in eq. (\ref{EqForOmega}). Requiring $\ddot{\Omega}^2\ll\dot{\Omega}^4$ is equivalent to $H\ll M$, and the condition $|\dddot{\Omega}\dot{\Omega}|\ll\dot{\Omega}^4$ is the same as $|dH/dt|\ll M^2$. Thus, the solution (\ref{VinFRW}) is accurate in the post-inflationary Universe.

Consider now the magnetic condensate. For concreteness, choose the Ansatz  
\begin{equation}
V_1=v(\eta)e^{-i\Omega(\eta)} \;, ~~~ V_2=iV_1 
\end{equation}
and $V_0=V_3=0$. The second of eqs. (\ref{TwoEqs}) remains unchanged and the first modifies to 
\begin{equation}\label{TwoEqs2}
\ddot{v}-v\dot{\Omega}^2+vM^2a^2-2v^3\gamma^2=0 \;.
\end{equation}
The relation (\ref{v(omega)}) between the magnitude and phase stays the same; plugging it to eq. (\ref{TwoEqs2}) gives
\begin{equation}\label{EqForOmega2}
-\dfrac{1}{2}\dddot{\Omega}\dot{\Omega}+\dfrac{3}{4}\ddot{\Omega}^2-\dot{\Omega}^4 +M^2a^2\dot{\Omega}^2-2\gamma^2v_*^2\dot{\Omega}=0 \; .
\end{equation}
Let us neglect the first two terms in eq. (\ref{EqForOmega2}). Make the following substitution for $\dot{\Omega}$:
\begin{equation}\label{OmegaA}
\dot{\Omega}=a\omega=aM\left(1-\dfrac{\epsilon}{2}\right) \; , ~~~ \epsilon\ll 1 \; .
\end{equation}
Plugging this to eq. (\ref{EqForOmega2}) yields
\begin{equation}
v_*^2=\dfrac{\epsilon a^3M^3}{2\gamma^2} \; .
\end{equation} 
Since $v_*$ must be independent of $\eta$, we have
\begin{equation}\label{FRW_eps}
\epsilon=\epsilon_*\dfrac{a_*^3}{a^3} \; .
\end{equation}
This means that as the condensate evolves, its angular velocity is pushed towards the non-relativistic limit $\omega\approx M$ in the FRW background. From eq. (\ref{OmegaA}) we have
\begin{equation}
e^{-i\Omega}=e^{-iMt}e^{\frac{iM\epsilon_*}{2}\int\frac{a_*^3dt}{a(t)^3}} \; .
\end{equation}
When $\epsilon\ll 1$, $a_*\ll a$, we can neglect the second phase in this expression. Thus, we arrive at
\begin{equation}
V_1(t)=V_{1*}\sqrt{\frac{a_*}{a(t)}} e^{-iMt} \; , ~~~ V_2(t)=iV_1(t) \;.
\end{equation}
This matches eqs. (\ref{VC_V_FRW}), (\ref{VC_V_FRW2}) in the non-relativistic limit.

The approximation we made in neglecting the first two terms in eq. (\ref{EqForOmega2}) is valid as long as
\begin{equation}
\begin{split}
& |\ddot{\Omega}|\ll |\dot{\Omega}|^2 ~~ \Rightarrow ~~ |\dot{a}\omega|\ll a^2\omega^2 \; , \\
&  |a\dot{\omega}|\ll a^2\omega^2 ~~ \Rightarrow ~~ H\ll aM \; , ~~ \epsilon H\ll a M \; .
\end{split}
\end{equation}
If we take $a>a_*\equiv\bar{a}$ with $\bar{a}$ the moment of formation of the magnetic condensate, then these conditions are automatically satisfied. 

\bibliography{cvdm}
 
\end{document}